\documentclass[aps,preprint,showpacs]{revtex4-1}
\usepackage[dvips]{graphics}
\begin{document}
\title{Field theory of low energy excitations of a mixture of two species of pseudospin-$\frac{1}{2}$ Bose gases with interspecies spin-exchange}
\author{Li Ge}
\affiliation{State Key Laboratory of Surface Physics and  Department of Physics, Fudan University, Shanghai,
200433, China}
\author{Yu Shi}
\email{
Email: yushi@fudan.edu.cn}
\affiliation{State Key Laboratory of Surface Physics and  Department of Physics, Fudan University, Shanghai,
200433, China}

\begin{abstract}

We develop a low energy effective field theory of a mixture of two species of pseudospin-$\frac{1}{2}$ atoms with interspecies
spin-exchange, in addition to  density-density interaction. This approach  is beyond the single orbital-mode approximation.  In a wide parameter regime, it indicates the existence of the four elementary excitations, especially a gapped mode due to interspecies spin-exchange.  On the other hand,    the spectrum of the effective spin Hamiltonian yielded by the single mode approximation can be obtained  by  quantizing the homogeneous excitation, which is  spin excitation and is the long-wavelength limit of the gapped mode of elementary excitations. These low energy excitations can be experimentally measured by using Bragg spectroscopy.

\end{abstract}


\maketitle

\section{Introduction}

Elementary excitations or collective modes are key properties of Bose-Einstein condensation (BEC), and serve as probes of the ground states.  The Bogoliubov theory of elementary excitations of BEC  gives an elegant description of the Goldstone modes associated with the spontaneous breaking of $U(1)$ symmetry~\cite{wen,altland,stoof}. In recent years, BEC of ultra-cold dilute atomic gases has become one of the most active fields in  physics. Among the most interesting topics are BEC of spinor atomic gases~\cite{stoof,pethick,pitaevskii,leggett2,leggett},  for example,  spin-1 and pseudospin-$\frac{1}{2}$ gases~\cite{spinor1,spinor2,kuklov,al,li}, as well as spinless mixtures~\cite{mix1,mix2}. As an extension of this topic, it is interesting to study spinor mixtures with interspecies spin exchange.    It has been theoretically found that
a mixture of two distinct species of pseudospin-$\frac{1}{2}$ atoms with interspecies spin-exchange interaction exhibits interesting features beyond both spinor gases and a mixture of spinless gases, especially, in a broad parameter regime, the ground state is entangled between the two species, rather than BEC of individual species~\cite{shi0,shi,shi2}. Also, the approach based on single orbital-mode approximation has  revealed interesting properties of quantum phase transition and many-particle quantum entanglement~\cite{shi,shi2,shi3,wu}.

We expect our work motivates more investigations along this line of research.
Spin-exchange scattering between distinguishable atoms has been less studied, perhaps because of incomplete information on inter-atomic potential.  However, we note that interspecies spin-exchange interaction can be significant.
There are  calculations indicating significant spin-exchange scattering lengths between distinguishable atoms~\cite{dalgarno}. Spin-changing scattering between distinguishable atoms has indeed been  observed~\cite{mudrich}. Experiments on multi-component Bose gases often had atom loss due to spin exchanges~\cite{spinor2,mix2}.
Significant differences between singlet and triplet scattering lengths have been observed in $^{41}$K-$^{87}$Rb, $^{40}K$-$^{87}$Rb and $^6$Li($^7$Li)-$^{23}$Na mixtures~\cite{ferrari}, implying significant interspecies spin exchanges. It is feasible to experimentally realize the systems studied here. One may use, for example,  $^{85}Rb$ and $^{87}Rb$, or $^{41}K$ and $^{87}Rb$, as the two species, and $|F=1,m_F=2\rangle$ and $|F=1,m_F=1\rangle$ as the two pseudospin states~\cite{shi2}.

In this paper, we treat a mixture of two distinct species of pseudospin-$\frac{1}{2}$ Bose gases by using a field theory approach beyond single orbital-mode approximation. From the point of view of field theory, there are four fields, as there are two species of atoms while each atom has two relevant spin states. We shall use the path integral formalism to develop a Bogoliubov-like mean field theory, in which each field has a specific value in the ground state. Excitations are then calculated as small deviations of the fields from those in the ground state.

Previously, elementary excitations in such a mixture have been studied as fluctuations of the single-particle orbital wave functions, and it has been restricted to a special parameter point, in which the many-body ground state is the so-called entangled BEC~\cite{shi3}. On the other hand, when the atoms are all condensed in the same  orbital wave functions,  there are spin excitations described by the effective spin Hamiltonian~\cite{shi}.  In the approach here, the spin excitations are  obtained as due to spin flipping of the fields that remain spatially homogeneous,  while  the elementary excitations are plane-wave-like excitations of the phases of the fields.  The former is the  long-wavelength limit of the gapped mode among the elementary excitation.   Furthermore, the  low energy effective theory gives the excitation spectrum of the effective spin Hamiltonian that is obtained under single orbital-mode approach.

\section{The model \label{model}}

Consider a dilute gas of two species of bosonic
atoms, the number of atoms in each species is conserved. Each atom possesses an internal degree of freedom represented as a pseudospin
with $z$-component basis states $|\uparrow\rangle$ and $|\downarrow\rangle$, and can transit between the two.
This system is described by four interacting boson fields, with the Lagrangian density
\begin{equation}
\mathcal{L}=\sum_{\alpha\sigma}{i \psi_{\alpha\sigma}^\dagger
\partial_t\psi_{\alpha\sigma}}-(\mathcal{H}-\sum_{\alpha\sigma}{\mu_{\alpha\sigma}
\psi_{\alpha\sigma}^\dagger
\psi_{\alpha\sigma}})
\end{equation}
where $\alpha=a, b$ represent the two species and $\sigma=\uparrow, \downarrow$
represents the two basis states of pseudospin-$\frac{1}{2}$,  $\psi_{\alpha\sigma}=\psi_{\alpha\sigma}(\mathbf{x})$ and  $\mu_{\alpha\sigma}$  are  the field  and the  chemical potential corresponding to the atoms of species $\alpha$ with pseudospin $\sigma$, respectively, $\mathcal{H}$ is the Hamiltonian density
\begin{equation}
\begin{array}{rl}
\displaystyle
\mathcal{H}=&
\displaystyle  \sum_{\alpha\sigma} \psi_{\alpha\sigma}^\dagger (-\frac{1}{2m_\alpha}\nabla^2 +   V) \psi_{\alpha\sigma} + \frac{1}{2}\sum_{\alpha\sigma\sigma^{'}}{
g^{(\alpha\alpha)}_{\sigma\sigma^{'}}
|\psi_{\alpha\sigma}|^{2}|\psi_{\alpha\sigma^{'}}|^{2}} \\
\displaystyle
&
\displaystyle +\sum_{\sigma\sigma^{'}}{g^{(ab)}_{\sigma\sigma^{'}}|\psi_{a\sigma}|^{2}
|\psi_{b\sigma^{'}}|^{2}}+
g_{e}(\psi_{a\uparrow}^{\dagger}
\psi_{a\downarrow}\psi_{b\downarrow}^{\dagger}\psi_{b\uparrow}+
\psi_{a\downarrow}^{\dagger}\psi_{a\uparrow}
\psi_{b\uparrow}^{\dagger}\psi_{b\downarrow}),
\end{array}
\end{equation}
where $V=V(x)$ is the external potential, $g^{(\alpha\alpha)}_{\sigma\sigma^{'}}$, $g^{(ab)}_{\sigma\sigma^{'}}$ and $g_{e}$ are the
interaction strengths for intraspecies scattering, interspecies scattering without spin exchange, and interspecies spin-exchange scattering respectively,  proportional to the corresponding scattering length.  For pseudospin-$\frac{1}{2}$ atoms, intraspecies scattering strengths with and without spin-exchange are the same~\cite{leggett}.

For simplicity, we set $V=0$, and also assume
$g^{(\alpha\alpha)}_{\sigma\overline{\sigma}}
=g^{(\alpha\alpha)}_{\sigma\sigma}=g_{\alpha}$, $g^{(ab)}_{\uparrow\uparrow}=g^{(ab)}_{\downarrow\downarrow}=g_s$ and  $g^{(ab)}_{\uparrow\downarrow}=g^{(ab)}_{\downarrow\uparrow}=g_d$
such that $\mu_{\alpha\uparrow}=\mu_{\alpha\downarrow}=\mu_{\alpha}$~\cite{shi2}.
We can define  $\Psi_{\alpha}(x) = (\psi_{\uparrow}(x),\psi_{\downarrow}(x))^T$, the spin density operator  ${\cal S}_{\alpha i}(x)=\psi_{\alpha\sigma}^{\dagger}(x)s^{i}_{\sigma\sigma'}
\psi_{\alpha\sigma'}(x)=\Psi^{\dagger}_{\alpha}{s}^{i}\Psi_{\alpha}$, $(i=x,y,z)$, where  $s^{i}=\tau_i/2$ is the single spin operator,  $\tau_i$ being the Pauli matrix. Then the
Hamiltonian density can be written as
\begin{equation}
\begin{array}{rl}
\displaystyle
{\cal H} =  &
\displaystyle \sum_{\alpha} \Psi_{\alpha}^\dagger (-\frac{1}{2m_\alpha}\nabla^2 +   U ) \Psi_{\alpha} +
\frac{g_a}{2}|\Psi_a|^4+\frac{g_b}{2}|\Psi_b|^4
+\frac{g_{ab}}{2}|\Psi_a|^2|\Psi_b|^2
\\
\displaystyle &
\displaystyle
+2g_{e}({\cal S}_{ax}{\cal S}_{bx}+ {\cal S}_{ay}{\cal S}_{by})+2g_z {\cal S}_{az}{\cal S}_{bz},
\end{array}
\end{equation}
where $g_{ab} \equiv g_s+g_d$, $g_z \equiv g_s-g_d$.

If $g_e=0$, the system is a mixture without interspecies spin exchange, equivalent to a mixture of four scalar Bose gases. Note that intraspecies spin exchange does not change the particle number occupying each pseudospin state. The Hamiltonian would possess a symmetry of $U(1)\times U(1)\times U(1)\times U(1)$,
corresponding to particle number conservation of all the four fields. With $g_e>0$,  the symmetry is lowered to $U(1)\times U(1)\times U(1)$,
corresponding to the conservations of  $N_a$, $N_b$ as well as $N_{a\uparrow}-N_{a\downarrow}+N_{b\uparrow}-N_{b\downarrow}=2 S_z$~\cite{shi3}.

\section{Effective lagrangian of low energy excitations  \label{theory} }

We consider the parameter regime $g_e > g_z$, with all the other parameters fixed. Other parameter regimes are studied elsewhere.  As a mean field theory, we suppose that
in the ground state, each of the four fields has a definite value $\psi_{\alpha \sigma}=\psi_{\alpha \sigma}^0 e^{i\Phi_{\alpha \sigma}^0}$. Then the spin exchange term
becomes $2g_e\psi_{a \uparrow}^0\psi_{a \downarrow}^0\psi_{b \uparrow}^0\psi_{b \downarrow}^0 \cos( \Phi_{a\uparrow}^0-\Phi_{a\downarrow}^0-\Phi_{b\uparrow}^0+\Phi_{b\downarrow}^0)$. Minimizing the potential part of the Lagrangian requires $\psi_{a\uparrow}^0=\psi_{a\downarrow}^0=\sqrt{\rho_a/2}$,
$\psi_{b\uparrow}^0=\psi_{b\downarrow}^0=\sqrt{\rho_b/2}$ and $\Phi_{a\uparrow}^0-\Phi_{a\downarrow}^0-\Phi_{b\uparrow}^0+\Phi_{b\downarrow}^0=\pi$. We can arbitrarily choose the phases of the four fields under the above constraint to describe  a ground state, other choices are equivalent in the sense of spontaneous symmetry breaking. Therefore in the ground state, $\psi_{a\uparrow}=\psi_{a\downarrow}=\psi_a^0=\sqrt{\rho_a/2}$,
$\psi_{b\uparrow}=-\psi_{b\downarrow}=\psi_b^0=\sqrt{\rho_b/2}$, where $\rho_{\alpha}=N_a/\Omega$ is
the number density of species $\alpha$, with $\Omega$ being the volume of the system. The  chemical potential is  evaluated to be  $\mu_{\alpha}=g_{\alpha} \rho_{\alpha} + \frac{1}{2}(g_{ab}-g_e)\rho_{\beta},$ where $\beta\neq \alpha$.

We now study the elementary excitations. With a deviation from the mean field value, each field can be written as $ \psi_{\alpha\sigma}=
\psi_\alpha^0(1+\zeta_{\alpha\sigma})e^{i\Phi_{\alpha\sigma}}$, where $\zeta_{\alpha\sigma}$ is a small quantity. Therefore
\begin{eqnarray}
\displaystyle
&\mathcal{L}=
\displaystyle
\sum_{\alpha\sigma}
[-\frac{\rho_\alpha}{2}\partial_t\Phi_{\alpha\sigma}
 -\frac{\rho_\alpha}{4m_\alpha}(\nabla{\Phi_{\alpha\sigma}})^2-
 \rho_\alpha\zeta_{\alpha\sigma}\partial_t\Phi_{\alpha\sigma}] \nonumber\\
\displaystyle
&
\displaystyle +\frac{g_e}{2}\rho_a\rho_b\cos{(\Phi_{a\uparrow}-\Phi_{a\downarrow}-\Phi_{b\uparrow}+\Phi_{b\downarrow})}
 -\frac{1}{2}
 \sum_{\alpha\beta,\sigma\sigma^{'}}
 \mathcal{Q}_{\alpha\sigma,\beta\sigma^{'}}
 \zeta_{\alpha\sigma}\zeta_{\beta\sigma^{'}},
 \end{eqnarray}
where  higher order and  constant terms have been neglected.
\begin{equation}
\mathcal{Q}=\left(
  \begin{array}{cc}
  \mathbf{H_a} & \mathbf{H_{ab}} \\
  \mathbf{H_{ab}} & \mathbf{H_b} \\
  \end{array}
  \right),
\end{equation}
is a $4\times 4$ matrix, with
\begin{equation}
 \mathbf{H}_\alpha=\rho_\alpha\left(
  \begin{array}{cc}
  g_\alpha\rho_\alpha+\frac{1}{2}g_e\rho_\beta, & g_\alpha\rho_\alpha-\frac{1}{2}g_e\rho_\beta \\
  g_\alpha\rho_\alpha-\frac{1}{2}g_e\rho_\beta, & g_\alpha\rho_\alpha+\frac{1}{2}g_e\rho_\beta \\
  \end{array}
  \right),
\end{equation} where $\alpha\neq{\beta}$,
\begin{equation}
\mathbf{H_{ab}}=\rho_a\rho_b\left(
  \begin{array}{cc}
  g_s-\frac{1}{2}g_e & g_d-\frac{1}{2}g_e \\
  g_d-\frac{1}{2}g_e & g_s-\frac{1}{2}g_e \\
  \end{array}
  \right).
\end{equation}

Now consider the vacuum persistence amplitude
$Z=\prod_{\alpha,\sigma} \int \mathcal{D}\Phi_{\alpha\sigma}
  \mathcal{D}\zeta_{\alpha\sigma} e^{i\int dt \int d^3x \mathcal{L}}$, from which  we obtain an effective Lagrangian as a function of $\Phi$ only, after
dropping the total time derivative of
  $\Phi$, which does not affect the equation of motion, and integrating over $\zeta$,
  \begin{eqnarray}
\displaystyle
  &
\displaystyle  \mathcal{L}_{eff}=
\frac{1}{2}\sum_{\alpha\beta,\sigma\sigma^{'}}{\rho_\alpha\rho_{\beta}(\partial_t\Phi_{\alpha\sigma})}
  (\mathcal{Q})^{-1}_{\alpha\sigma,\beta\sigma^{'}}
  (\partial_t\Phi_{\beta\sigma^{'}}) - \sum_{\alpha\sigma}
  {\frac{\rho_\alpha}{4m_\alpha}(\nabla{\Phi_{\alpha\sigma}})^2} \nonumber\\
\displaystyle
  &
\displaystyle -\frac{g_e}{2}\rho_a\rho_b \cos(\Phi_{a\uparrow}-\Phi_{a\downarrow}-\Phi_{b\uparrow}+\Phi_{b\downarrow})
  \end{eqnarray}
In deriving this formula, we neglect the $(\nabla{\zeta})^2$ and $\zeta^3$, $\zeta^4$ terms since only low energy dynamics is concerned. This Lagrangian has a cosine term
similar that in the sine-Gordon model. In $1+1D$, this term leads to a solution of topological soliton, which has very nontrivial contribution to the phase diagram, as discussed elsewhere. However, in this paper we focus on the low energy limit in  $3+1D$ case, in which the fluctuation of $\Phi_{a\uparrow}-\Phi_{a\downarrow}-\Phi_{b\uparrow}+\Phi_{b\downarrow}$ is largely suppressed and we can make the approximation $\cos{x} \approx 1-x^2/2$.

The conjugate relation between the phase
$\Phi_{\alpha\sigma}$ and particle number $N_{\alpha\sigma}$, the conservation of  $N_\alpha=\sum_{\sigma} N_{\alpha\sigma}$ and the fact that
the mass term  is proportional to $(\Phi_{a\uparrow}-\Phi_{a\downarrow}-\Phi_{b\uparrow}+\Phi_{b\downarrow})^2$
suggest a transformation
 \begin{equation}
 \Gamma = U \mathbf {\Phi}
 \end{equation}
 where $\Gamma \equiv (\gamma_1,\gamma_2,\gamma_3,\gamma_4)^{T}$, $\mathbf{\Phi} \equiv (\Phi_{a\uparrow},\Phi_{a\downarrow},\Phi_{b\uparrow},
 \Phi_{b\downarrow})^T$,
\begin{equation}
       U \equiv  \left(
           \begin{array}{cccc}
             \frac{1}{\sqrt{2}} & \frac{1}{\sqrt{2}} & 0 & 0 \\
             0 & 0 & \frac{1}{\sqrt{2}} & \frac{1}{\sqrt{2}} \\
             \frac{1}{2} & -\frac{1}{2} & \frac{1}{2} & -\frac{1}{2} \\
             \frac{1}{2} & -\frac{1}{2} & -\frac{1}{2} & \frac{1}{2} \\
           \end{array}
         \right),
  \end{equation}
  which is orthogonal, i.e.
  ${U}^{-1}={U}^T$. Then the effective
  Lagrangian can be rewritten as
  \begin{equation}
  \mathcal{L}_{eff}=\frac{1}{2} (\partial_t\Gamma^T) {A}^{-1}(\partial_t\Gamma)
  -\frac{1}{2}(\nabla{\Gamma}^T) M^{-1}(\nabla{\Gamma})
  -\Gamma^T G \Gamma
  \end{equation}
  where ${A}^{-1} = {U}{D}^T{\cal Q}^{-1} {D}{U}^{-1} $ is symmetric,
  \begin{equation}
 {D}=D^T=\left(
                                        \begin{array}{cccc}
                                          \rho_a & 0 & 0 & 0 \\
                                          0 & \rho_a & 0 & 0 \\
                                          0 & 0 & \rho_b & 0 \\
                                          0 & 0 & 0 & \rho_b \\
                                        \end{array}
                                      \right).
  \end{equation}
 Hence
  \begin{equation}
 {A}=UD^{-1}{\cal Q}D^{-1}U^T =\left(
                                        \begin{array}{cccc}
                                          2g_a & g_{ab}-g_e & 0 & 0 \\
                                          g_{ab}-g_e & 2g_b & 0 & 0 \\
                                          0 & 0 & g_e\eta_{+}+g_z & g_e\eta_{-} \\
                                          0 & 0 & g_e\eta_{-} & g_e\eta_{+}-g_z \\
                                        \end{array}
                                      \right),
  \end{equation}
  where $\eta_{\pm } =\frac{1}{2}(\frac{\rho_b}{\rho_a} \pm \frac{\rho_a}{\rho_b})$,
  \begin{equation}
 {M}^{-1}=\frac{1}{2} \left(
               \begin{array}{cccc}
                 \frac{\rho_a}{m_a} & 0 & 0 & 0 \\
                 0 & \frac{\rho_b}{m_b} & 0 & 0 \\
                 0 & 0 & \xi_+ & \xi_- \\
                 0 & 0 & \xi_- & \xi_+ \\
               \end{array}
             \right),
  \end{equation}
 where $\xi_{\pm } =\frac{1}{2}(\frac{\rho_a}{m_a}\pm \frac{\rho_b}{m_b})$,
 \begin{equation}
 G= \left(
               \begin{array}{cccc}
                 0 & 0 & 0 & 0 \\
                 0 & 0 & 0 & 0 \\
                 0 & 0 & 0 & 0 \\
                 0 & 0 & 0 & g_e\rho_a\rho_b \\
               \end{array}
             \right).
  \end{equation}

 From Euler-Lagrange equation
 \begin{equation}
 \partial_t(\frac{\partial {\cal L}_{eff}}{\partial (\partial_t \Gamma^T)})
+\nabla (\frac{\partial {\cal L}_{eff}}{\partial (\nabla \Gamma^T)})
-\frac{\partial{\cal L}_{eff}}{\partial  \Gamma^T} = 0,
\end{equation}
we obtain the equation of motion of $\Gamma$,
\begin{equation}
 \partial_t^2  \Gamma -
{A} {M}^{-1} \nabla^2   \Gamma + 2{A}{G} \Gamma  = 0,
\end{equation}

\section{Elementary excitations  \label{elementary} }

For elementary excitations, as characterized by frequency $\omega$ and wave vector $\mathbf{k}$,  we seek solutions of the form of
\begin{equation}
\Gamma=\Gamma_0\exp [-i(\omega t-\mathbf{k}\cdot \mathbf{r})],
\end{equation}
where $\Gamma_0$ is position independent.
Hence we obtain
\begin{equation}
 (-\omega^2 + k^2 A M^{-1} + 2{A}{G}) \Gamma_0  = 0.
\end{equation}

The secular equation $det(-\omega^2 + k^2 A M^{-1} + 2{A}{G})=0$ gives
\begin{equation}
\left|
\begin{array}{cccc}
-\omega^2+\frac{g_a\rho_ak^2}{m_a}& \frac{(g_{ab}-g_e)\rho_bk^2}{2m_b}&0&0\\
\frac{(g_{ab}-g_e)\rho_ak^2}{2m_a}&-\omega^2+\frac{g_b\rho_b k^2}{m_b}& 0 &0\\0&0&  -\omega^2 + \frac{k^2[(g_e f_1 +g_z\xi_+)}{2}&
 \frac{k^2(g_e f_2 + g_z\xi_-)}{2} +2g_e^2\eta_- \rho_a\rho_b \\ 0&0 &
 \frac{k^2(g_e f_2-g_z\xi_-)}{2} & -\omega^2 + \frac{k^2(g_e f_1- g_z\xi_+)}{2}+2g_e(g_e\eta_+-g_z) \rho_a\rho_b
                               \end{array}
                             \right|=0
                             \end{equation}
 where
$f_1 \equiv \eta_+ \xi_+ + \eta_-\xi_-$, $f_2\equiv \eta_+\xi_- + \eta_-\xi_+$.

It is found that the four excitations are given by
\begin{equation}
\omega^2_{I,II}=\frac{k^2}{2}\bigg[\frac{g_a\rho_a}{m_a}+\frac{g_b\rho_b}{m_b}
\mp \sqrt{(\frac{g_a\rho_a}{m_a}-\frac{g_b\rho_b}{m_b})^2+
\frac{(g_{ab}-g_e)^2\rho_a\rho_b}{m_am_b}}\bigg],  \label{excitation1}
\end{equation}
\begin{equation}
\omega^2_{III,IV}= \frac{1}{2}\bigg[ Bk^2
+\Delta^2
\mp \sqrt{Ck^4+Dk^2+\Delta^4 }\bigg],  \label{excitation2}
\end{equation}
where
\begin{equation}
\Delta^2 =  |g_e^2 (\frac{\rho_b}{\rho_a}+\frac{\rho_a}{\rho_b})-2g_eg_z|\rho_a\rho_b, \label{gap}
\end{equation}
$B\equiv \frac{g_e}{2}(\frac{\rho_b}{m_a}+\frac{\rho_a}{m_b})$,
$C\equiv \frac{g_e^2}{4}(\frac{\rho_b}{m_a}-\frac{\rho_a}{m_b})^2+g_z^2\frac{\rho_a\rho_b}{m_am_b}$,
$D\equiv g_e\rho_a\rho_b[g_e^2 (\frac{\rho_b}{m_a}-\frac{\rho_a}{m_b})(\frac{\rho_b}
{\rho_a}-\frac{\rho_a}{\rho_b})-2g_eg_z(\frac{\rho_b}{m_a}+\frac{\rho_a}{m_b})+2g_z^2(\frac{\rho_a}{m_a}+\frac{\rho_b}{m_b})]$.

It can be seen that $\omega_{IV}$ has a gap $\Delta$, due to the nonvanishing $g_e$,   while the other three excitations, as Goldstone modes, are gapless.
As $k\rightarrow 0$,
\begin{equation}
\omega^2_{III} \approx \frac{1}{2}(B-\frac{D}{2\Delta^2}) k^2 -\frac{C}{4\Delta^2} k^4,
\end{equation}
\begin{equation}
\omega^2_{IV} \approx \Delta^2 +\frac{1}{2}(B+\frac{D}{2\Delta^2}) k^2 +\frac{C}{4\Delta^2} k^4,
\end{equation}
When $\rho_a=\rho_b=\rho$, we have $B=\frac{1}{2}g_e(\frac{1}{m_a}+\frac{1}{m_b})\rho$, $D=-2g_eg_z(g_e-g_z)(\frac{1}{m_a}+\frac{1}{m_b})\rho^3$,  $\Delta^2=2g_e(g_e-g_z)\rho^2$.
Note that all our calculations are under the presumption that $g_e>g_z>0$.

$\omega^2$ may be negative in some cases, which means that the mean-field ground state with $\psi_{a\uparrow}=\psi_{a\downarrow}=\sqrt{\rho_a/2}$, $\psi_{b\uparrow}=-\psi_{b\downarrow}=\sqrt{\rho_b/2}$ is unstable and a phase transition occurs.
From the secular equation we see that $\omega^2 \geq 0$ is satisfied for any $k$ if and only if the matrix ${A}$ is positive definite, as the matrix ${M^{-1}}$ is positive definite while $G$ is semi positive-definite. This means $g_a>0$, $g_b>0$, $4g_ag_b>(g_{ab}-g_e)^2$ and $g_e>g_z$. The first three conditions can be naturally satisfied.  If $g_e<g_z $, we have $B^2-C=(g_e^2-g_z^2)\frac{\rho_a\rho_b}{m_am_b}<0$ and $2B\Delta^2-D=2(g_e^2-g_z^2)(\frac{\rho_a}{m_a}+\frac{\rho_b}{m_b})<0$, then $\omega^2_{III}$ becomes negative for any $k$ and fluctuations will destroy the mean-field ground state to form a new phase.

The parameter point of $g_e=g_z$ is a point of quantum phase transition. The gap $\Delta$ calculated above vanishes at this point, signalling the inappropriateness of the present mean field theory for this phase.  Indeed the phase at $g_e=g_z$ is  the so-called entangle BEC discussed previously by using the single orbital-mode approximation, in which the two species are maximally entangled in their collective spins, and BEC occurs in an interspecies two-particle singlet state. At  $g_e=g_z$, the gap calculated in a single orbital-mode approximation does not vanish, but is maximal on the contrary~\cite{shi3,wu}. An appropriate  field theory for this phase is under development.

We can also obtain the correlation function $<T[\gamma_\mu(t,x)\gamma_\nu(0,0)]>$,
 $\mu,\nu=1,2,3,4$. What interests us most is $<T[\gamma_4(t,x)\gamma_4(0,0)]>$. In
 momentum space:
 $$G_4(k,\omega)=-i<[\gamma_4(k,\omega)\gamma_4(-k,-\omega)]>
 =\frac{g_e\eta_{+}-g_z}{\omega^2-\omega_{IV}^2+i0^{+}}$$
 By neglecting the $k^4$ term in $\omega_{IV}$ we obtain
 \begin{eqnarray}
 iG_4(x,t)=i\int{\frac{d^3kd\omega}{(2\pi)^4}\frac{g_e\eta_{+}-g_z}
 {\omega^2-\omega_{IV}^2+i0^{+}}e^{i(\mathbf{k}\cdot\mathbf{x} -\omega t)}}\nonumber\\
 =-\frac{g_e\eta_{+}-g_z}{4\pi^2vr}\partial_rK_0(v\sqrt{(r^2-v^2t^2)}\Delta) \label{green}
 \end{eqnarray}
 where $r=|\mathbf{x}|$, $v=\sqrt{\frac{B}{2}+\frac{D}{4\Delta^2}}$, and $K_0(z)$ is the
 Modified Bessel Function of the Second Kind which has the following asymptotic behavior:
 $$K_0(z)=\left\{ \begin{array}{rl}
 -\ln{z} & \ z\ll{1} \\
 \sqrt{\frac{\pi}{2z}}e^{-z} &  z\gg{1}
 \end{array} \right.$$

 From $G_4(x,t)$, we can also obtain the correlation function of spin-exchange operator ${\cal S}_e \equiv \psi_{a\downarrow}^\dagger \psi_{a\uparrow} \psi_{b\uparrow}^\dagger \psi_{b\downarrow}$,
\begin{equation}
 \langle {\cal S}_e^{\dagger}(\mathbf{x},0) {\cal S}_e(0,0) \rangle \propto \langle e^{-i\gamma_4(x,0)}
 e^{i\gamma_4(0,0)} \rangle =e^{iG_4(\mathbf{x},0)}e^{-iG_4(\mathbf{l},0)},
 \end{equation}
 where $\mathbf{l}$ is a vector of short-range cut-off length. According to (\ref{green}),
 With $g_e >g_z$,  $\langle {\cal S}_e^{\dagger}(\mathbf{x},0){\cal S}_e(0,0) \rangle$  decreases with the increase of $r$.

\section{Homogeneous excitations \label{homogeneous} }

We now consider a homogeneous excitation $\Gamma(x,t)=\Gamma(t)$, i.e. fluctuations purely caused by spin flipping while the orbital wave functions remain homogeneous.
In this case the effective Lagrangian is
$L_{homo}=\int{d^3x\mathcal{L}_{homo}}
 =\Omega[\frac{1}{2}(\partial_t\Gamma)^T A^{-1}(\partial_t\Gamma)
 -g_e\rho_a\rho_b\gamma_4^2]$. The canonical momentum conjugate with $\gamma_{\mu}$ is
 \begin{equation}
 p_\mu=\frac{\partial L}{\partial\dot{\gamma_\mu}}
 =\Omega A^{-1}_{\mu\nu}\gamma_\nu,
  \end{equation}
  ($\mu,\nu=1,2,3,4$).
  The effective Hamiltonian is thus
  \begin{eqnarray}
 H_{homo}= p_\mu\dot{\gamma_\mu}-L=\frac{1}{2\Omega}p_\mu\mathbf{A}_{\mu\nu}p_\nu
 +\Omega g_e\rho_a\rho_b\gamma_4^{2}. \label{hamil} \end{eqnarray}

 Quantization of these excitations is done by imposing the commutation relation
 $[\gamma_\mu,p_\nu]=i\delta_{\mu\nu}$.
 Recalling  $\gamma_\mu = U_{\mu\rho} \Phi_\rho$ and that the $N_{\alpha\sigma}$ and $\Phi_{\alpha\sigma}$ are conjugated variables with
 $[N_{\alpha\sigma}, \Phi_{\beta\sigma^{'}}]=i\delta_{\alpha\beta,\sigma\sigma^{'}}$, we obtain
 \begin{equation}
p_{\nu} = - U_{\nu\rho} N_{\rho},
 \end{equation}
 where $N_\rho$ represents $N_1= N_{a\uparrow}$, $N_2=N_{a\downarrow},$ $N_3= N_{b\uparrow}$ and $N_4=N_{b\downarrow}$. That is,
\begin{eqnarray}
& p_1=-\frac{1}{\sqrt{2}}(N_{a\uparrow}+N_{a\downarrow})=-\frac{1}{\sqrt{2}}N_a, \nonumber \\
& p_2=-\frac{1}{\sqrt{2}}(N_{b\uparrow}+N_{b\downarrow})=-\frac{1}{\sqrt{2}}N_b, \nonumber \\
& p_3=-\frac{1}{2}(N_{a\uparrow}-N_{a\downarrow}+N_{b\uparrow}-N_{b\downarrow})=-S_z, \nonumber \\
& p_4=-\frac{1}{2}(N_{a\uparrow}-N_{a\downarrow}-N_{b\uparrow}+N_{b\downarrow}).
\end{eqnarray}

The effective Hamiltonian (\ref{hamil}) can be solved easily. It can be seen that $p_1$, $p_2$ and $p_3$ are conserved quantities, because  $N_a$, $N_b$ and $S_z$ are conserved. The  effective Hamiltonian can be rewritten as
\begin{equation}
H_{homo} =  \frac{1}{2\Omega}\sum_{i,j=1,2}{p_iA_{ij}p_j}
+ \frac{(g_e^2-g_z^2)}{2\Omega(g_e\eta_{+}-g_z)}p_3^{2}
+ \frac{g_e \eta_+-g_z}{2\Omega}{p_4'}^{2} + \Omega g_e\rho_a\rho_b\gamma_4^2,
\end{equation}
where $p_4' \equiv p_4 -\frac{g_e\eta_-}{g_e \eta_+-g_z} p_3$. $p_1=-N_a/\sqrt{2}$, $p_2=-N_b/\sqrt{2}$ and $p_3=-S_z$ are all conserved, while the part depending of $p_4'$ and $\gamma_4$ is like the Hamiltonian of a harmonic oscillator.
Therefore the spectrum of $H$ is
\begin{eqnarray}
 E_{homo}(p_1, p_2, p_3, n)&=&\frac{1}{2\Omega}\sum_{i,j=1,2}{p_iA_{ij}p_j}
 +\frac{(g_e^2-g_z^2)}{2\Omega(g_e\eta_{+}-g_z)}p_3^{2}+
 (n+\frac{1}{2})\Delta\nonumber\\
& =&E_0 +\frac{(g_e^2-g_z^2)}{2\Omega(g_e\eta_{+}-g_z)}S_z^{2}+
 (n+\frac{1}{2})\Delta
  \label{spec}
 \end{eqnarray}
 where
 \begin{equation}
 E_0 = \frac{1}{2\Omega}[g_aN_a^2+g_bN_b^2+(g_{ab}-g_e)N_aN_b]
 \end{equation}
is fixed, $\Delta$ is nothing but the energy gap in (\ref{gap}). In the ground state, $S_z=0$, $n=0$, thus  the energy is
\begin{eqnarray}
E(S_z=0,n=0)=E_0 +\frac{1}{2}\Delta,
 \end{eqnarray}
where $\Delta/2$ is the zero-point energy of homogeneous fluctuation (that is, the $k=0$ part of (\ref{excitation1}) and (\ref{excitation2})).  The excitation energy of the homogeneous excitation is
 \begin{equation}
 E_{homo}(S_z, n)-E(S_z=0, n=0)=\frac{(g_e^2-g_z^2)}{2\Omega(g_e\eta_{+}-g_z)}S_z^{2}
 +n\sqrt{2g_e(g_e\eta_{+}-g_z)\rho_a\rho_b}, \label{excite}
 \end{equation}
where $n=0,1,\cdots$.

It can be seen that the fourth elementary excitation, discussed in the last Section,  reduces to the homogeneous excitation as $k \rightarrow 0$.

\section{Single orbital-mode approximation \label{single} }

For the ground state of a Bose gas, usually the approximation of the single orbital mode works very well and is the common practice. For our system, this approximation means that only one orbital mode is contained in each field,  that is,  $\psi_{\alpha\sigma} \approx \alpha_{\sigma}  \phi_{\alpha\sigma}$, $\alpha_{\sigma}$  denotes the annihilation operator of the orbital mode function $\phi_{\alpha\sigma}$. Then the spin operator for species $\alpha$ is $\mathbf{S}_{\alpha} \equiv \int d^3 {\cal S}_{\alpha} (\mathbf{x}) = \alpha^{\dagger}_{\sigma}\mathbf{s}_{\sigma\sigma'} \alpha_{\sigma}$, and thus the Hamiltonian $H= \int d^3 x {\cal H}(x)$ becomes, up to a constant, \begin{equation}
H_s = K_e (S_{ax}S_{bx}+ S_{ay}S_{by}) + J_z S_{az}S_{bz},
\end{equation}
where $K_e$ and $J_z$ are effective parameters determined by the interaction strengths as well as single particle orbital wave functions and energies. Then the total spin  $S$ of the system  is conserved. In the the uniform case, $V=0$, it can be found that $K_e=2g_e/\Omega$, $J_z =2g_z/\Omega$, where $\Omega$ is the volume of the system.

Under the single orbital-mode approximation,  $\psi_{\alpha\sigma} \approx \alpha_{\sigma}  \phi_{\alpha\sigma}$,  the elementary excitation, that is, the fluctuated phase factor  $e^{i\Phi_{\alpha \sigma}}$   with a wave-like dependence on $\mathbf{x}$ and $t$, can only be attributed to the fluctuation of the orbital wave function  $\phi_{\alpha\sigma}$. This verifies the previous treatment of elementary excitation using the Gross-Pitaevskii-like equation governing the single-particle orbital wave functions. Only in the long-wavelength limit,  the gapped elementary excitations reduces the homogeneous excitation.

For a homogeneous excitation, the
phase factor  $e^{i\Phi_{\alpha \sigma} }$  is position independent, and be attributed to spin degree of freedom.
Hence a homogeneous excitation is a spin excitation, with the orbital degree of freedom remaining the same as those in the ground state. As such, these excitations should be the same as those of the effective spin Hamiltonian $H_s$ obtained under single-orbital mode approximation.

As such, the energy spectrum of a homogeneous excitation (\ref{excite}) can be approximately equalized with the spectrum of $H_s$, that is
\begin{equation}
E_s=E_{homo}=\frac{(K_e-J_z)^2}{4(K_e\eta_{+}-J_z)}S_z^{2}
+\frac{1}{2}n\sqrt{2K_e(K_e\eta_{+}-J_z)N_aN_b}
 \end{equation}
which, for  $N_a=N_b=N$,   reduces to
\begin{equation}
E_s=\frac{1}{4}(K_e+J_z)S_z^{2}+\frac{1}{2}n\sqrt{2K_e(K_e-J_z)}N. \label{en}
 \end{equation}

\begin{figure}
\begin{center}
\includegraphics{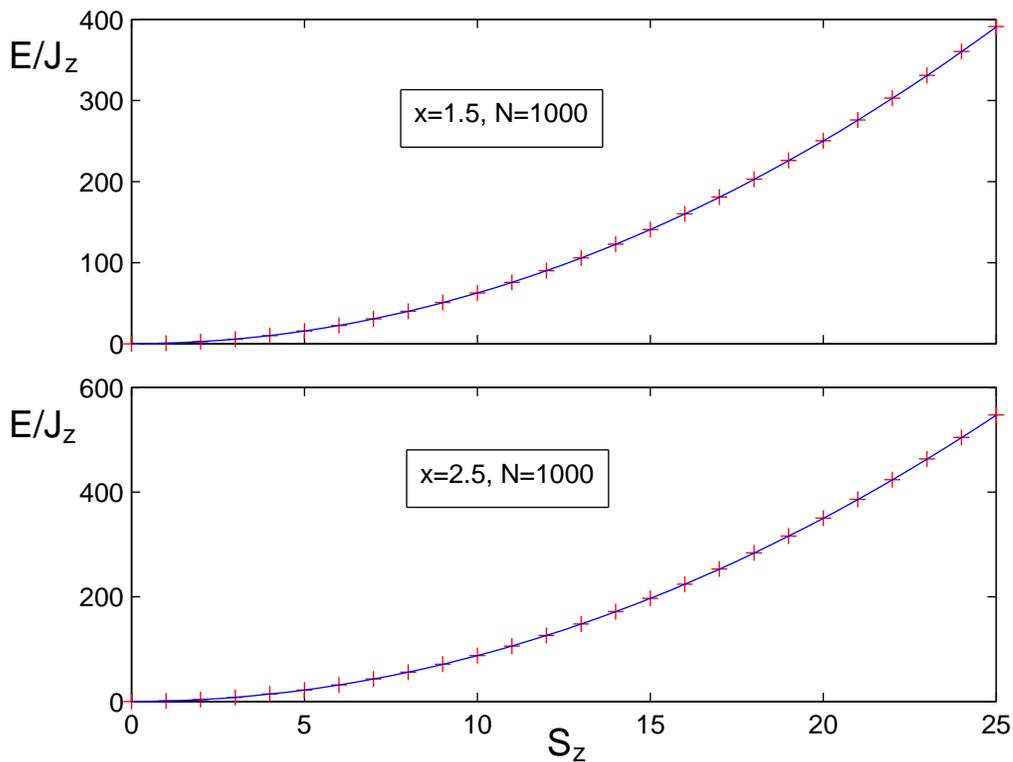}
\caption{ \label{fig1} $E$ as a function of $S_z$ with  $n=0$. $x=K_e/J_z$ and N is the particle
  number. The '+'s are the numerical solution of $H_s$, and the solid line is the plot of
$E=(K_e+J_z)S_z^2/4$. They fit extremely well.}
\end{center}
\end{figure}

\begin{figure}
\begin{center}
\includegraphics{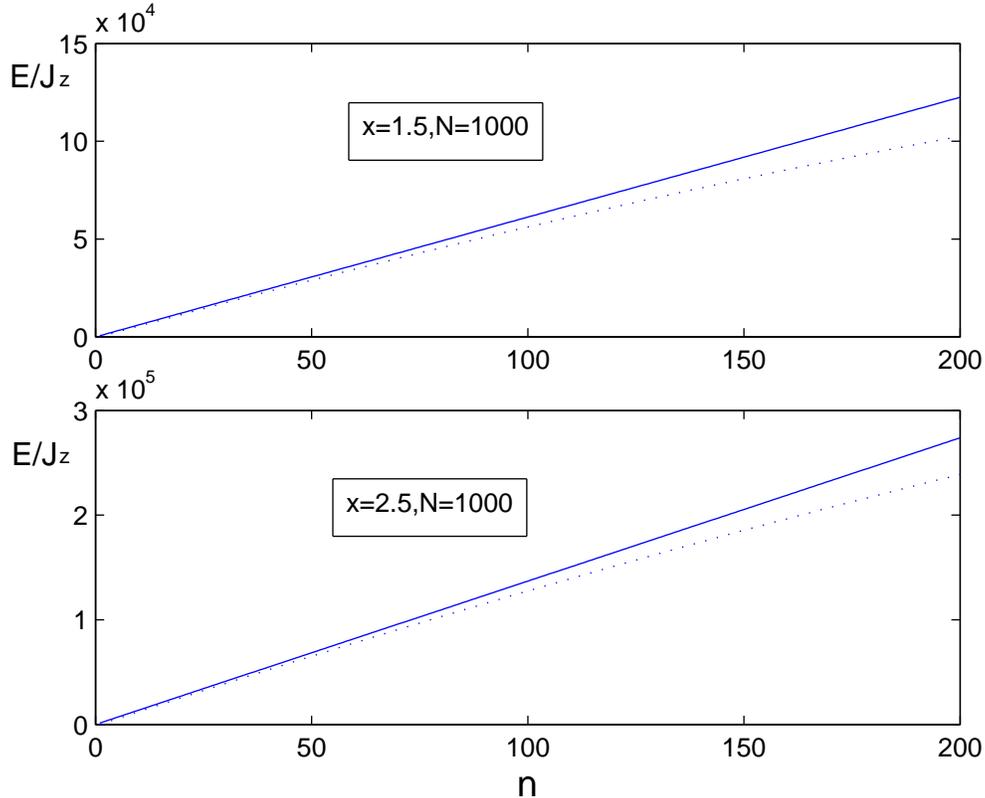}
\caption{\label{fig2}  $E$ as a function of $n$ with $S_z=0$. $x=K_e/J_z$ and N is the particle  number. The dashed line represents the numerical solution of $H_s$ and the solid line is the plot of
 $E=\frac{1}{2}n\sqrt{2K_e(K_e-J_z)}N$. Note that the low lying excited states correspond to small values of $n$, for which the low energy field theory and the single orbital mode approximation fit well.}
\end{center}
\end{figure}

We have numerically solved the effective spin Hamiltonian $H_s$  and compared the result with the the above expression of spectrum (\ref{en}). As shown in  Figs.~\ref{fig1} and \ref{fig2}, they fit very well for small $n$. For a Bose gas in absence of a magnetic field, $N$ is very large while $S_z$ is very small. Hence in (\ref{en}), unless $n=0$, the first term is much smaller than the second term. Therefore,
the low-lying states must be those with a certain small $S_z$ and with $n=0$.  This firmly indicates that our field theory and the single orbital-mode approximation fit very well for low energy excitations.

This result also confirms a previous perturbative treatment of the anisotropic coupling between the collective spins~\cite{shi2}. Both the unperturbed isotropic Hamiltonian and the anisotropic perturbation conserve $S_z$, hence the eigenstate is a superposition of  states $|S,S_z\rangle$ with a same value of $S_z$ and different values of $S$. The expansion coefficients turn out to be the ``wave functions'' of a harmonic oscillator in coordinate $S$, also giving the spectrum (\ref{en}). The total spin $S$  is indeed  equivalent to $\gamma_4\equiv (\Phi_{a\uparrow}-\Phi_{a\downarrow}-\Phi_{b\uparrow}
+\Phi_{b\downarrow})/2 $, as the angle between the collective spins of the two species is just  $2\gamma_4-\pi$~\cite{shi3}.

\section{Summary and discussion \label{summary} }

We have described the low energy excitations of a mixture of two species of pseudospin-$\frac{1}{2}$ Bose gases with interspecies spin exchanges, which entangles the two species of atoms when the system undergoes BEC.
From the point of view of quantum field theory, we have considered a four-component field with various interactions. We developed a low energy effective field theory, which can very well describe various low energy excitations in a unified framework. As an interesting generalization of the usual Bogoliubov theory for the present multicomponent Bose gas with spin degree of freedom, this theory gives four elementary excitations. The most interesting aspect is the gap in one of the four excitations. On the other hand, quantizing  homogeneous excitations yields the excitation spectrum which can be attributed to spin degree of freedom. Interestingly, this leads to an analytical solution of the effective spin Hamiltonian obtained under single orbital-mode approximation.

Notice that  in a realistic system in a trapping potential,  there is cut-off of  Goldstone modes due to the trap,  thus the low-energy excitations become discrete collective modes~\cite{pitaevskii,stringari}.

The elementary excitations or collective modes can be measured by using the Bragg spectroscopy, based on two-photon Bragg scattering~\cite{rmp}.
Especially, several modes coexisting at a given value of momentum transfer can be excited and measured~\cite{multi}.
For a trapped gas, the collective modes can also be measured by perturbing the trapping potential~\cite{jin}. Similarly, the excitations discussed in our work can be experimentally measured by using the above method.  The homogeneous excitations can also be measured, in a way similar to the measurement of collective modes in a trap, which are not plane waves~\cite{rmp}.

The gap in a collective mode is a feature nonexisting in the usual mixtures, where the particle number of each spin state is conserved~\cite{ferrari}. The nonvanishing value of $g_e$, which accounts for the gap, as well as $g_z$, which characterizes the difference of scattering lengths of like-spin and unlike-spin scattering processes,  both originate from the interspecies spin exchange interaction. Therefore $g_e$ and $g_z$ are roughly of the same order of magnitude. Many experiments have been carried out to measure this  interaction~\cite{ferrari}, indicating a considerably large value of scattering length, which is about  $100 a_B$,  where $a_B$ is the Bohr radius. Therefore we expect that experimentally this system can be realized and that the   gapped mode can be found.

\acknowledgments

This work was supported by the National Science Foundation of China (Grant Nos. 10875028 and 11074048) and the Ministry of Science and Technology of China (Grant No. 2009CB929204).

\end{document}